\documentclass[final,5p,times,twocolumn,authoryear]{elsarticle}
\usepackage{amssymb}
\usepackage{lipsum}
\usepackage{amsmath,amssymb,amsfonts}
\usepackage{algorithmic}
\usepackage{graphicx}
\usepackage{subcaption}
\usepackage{textcomp}
\usepackage{xcolor}
\usepackage{siunitx}  
\usepackage{svg}
\usepackage{bm}

\definecolor{patellar}{RGB}{202, 160,255}
\definecolor{femoral}{rgb}{0.565498, 0.84243 , 0.262877}
\definecolor{tibial}{RGB}{253, 170, 72}
\definecolor{meniscus}{RGB}{255, 255, 10}
\definecolor{fp}{rgb}{1, 0, 0}
\definecolor{fn}{RGB}{0, 100, 255}
\usepackage{booktabs}  
\usepackage{multirow}
\usepackage{hyperref}

\begin{document}
\begin{frontmatter}

\author[1]{Jan Nikolas Morshuis\corref{cor1}}
\ead{nikolas.morshuis@uni-tuebingen.de}
\author[1]{Matthias Hein}
\author[1,2]{Christian Baumgartner}

\affiliation[1]{organization={University of Tübingen},
            city={Tübingen},
            country={Germany}}

\affiliation[2]{organization={University of Lucerne},
             city={Lucerne},
             country={Switzerland}}

\title{Understanding Benefits and Pitfalls of Current Methods for the Segmentation of Undersampled MRI Data}

\begin{abstract}

MR imaging is a valuable diagnostic tool allowing to non-invasively visualize patient anatomy and pathology with high soft-tissue contrast. However, MRI acquisition is typically time-consuming, leading to patient discomfort and increased costs to the healthcare system. Recent years have seen substantial research effort into the development of methods that allow for accelerated MRI acquisition while still obtaining a reconstruction that appears similar to the fully-sampled MR image. 
However, for many applications a perfectly reconstructed MR image may not be necessary, particularly, when the primary goal is a downstream task such as segmentation. This has led to growing interest in methods that aim to perform segmentation directly on accelerated MRI data.
Despite recent advances, existing methods have largely been developed in isolation, without direct comparison to one another, often using separate or private datasets, and lacking unified evaluation standards. To date, no high-quality, comprehensive comparison of these methods exists, and the optimal strategy for segmenting accelerated MR data remains unknown.
This paper provides the first unified benchmark for the segmentation of undersampled MRI data comparing 7 approaches. A particular focus is placed on comparing \textit{one-stage approaches}, that combine reconstruction and segmentation into a unified model, with \textit{two-stage approaches}, that utilize established MRI reconstruction methods followed by a segmentation network. We test these methods on two MRI datasets that include multi-coil k-space data as well as a human-annotated segmentation ground-truth. We find that simple two-stage methods that consider data-consistency lead to the best segmentation scores, surpassing complex specialized methods that are developed specifically for this task.
\end{abstract}


\end{frontmatter}
\section{Introduction}

Magnetic Resonance Imaging (MRI) has proven to be a powerful modality to segment anatomical structures and pathologies due to its high soft-tissue contrast. However, a major disadvantage of MRI acquisition are the long scan times required. This does not only lead to high healthcare costs and longer waiting times but also to increased patient discomfort \citep{oztek2020practical} and potential artifacts in the MRI image due to patient movement \citep{zaitsev2015motion}. Faster MRI acquisition is therefore desirable and currently a major research topic \citep{heckel2024deeplearningacceleratedrobust}.

A widely used strategy to accelerate MRI acquisition is to reduce the amount of data collected, resulting in an undersampled k-space. This undersampling leads to gaps in the measurement data. Replacing these gaps with zeros, however, can lead to visible degradations of the MRI image, impacting it's usefulness for diagnostics. To address this issue, current reconstruction methods aim to estimate the fully-sampled MR image by utilizing some form of prior information. This prior can either be learned, as is the case for machine learning-based reconstruction approaches \citep{heckel2024deeplearningacceleratedrobust}, or it can be based on known properties of MR images like sparsity and smoothness constraints as it is done in compressed sensing methods \citep{donoho2006compressed,lustig2007sparse}. Utilizing prior information in addition to the available measurements typically yields reconstructions that more closely resemble the fully sampled ground truth MRI image.

A tradeoff exists between the quality of the reconstruction and the acceleration factor -- higher accelerations allow shorter measurement times but typically also lead to worse reconstructions. What acceleration factor to choose largely depends on the downstream task: For instance, \citep{radmanesh_muckley_murrell_lindsey_sriram_knoll_sodickson_lui_2022} have shown that even though an accurate reconstruction might not be possible for high accelerations, the reconstructed images can still be used for initial screening by radiologists, who can decide which patients need further examination. Therefore, the acceleration factor of the MRI can be increased significantly if the purpose of the acquisition is not the creation of accurate reconstructions but rather the downstream task of initial screening. Similarly, the ultimate purpose of an MRI acquisition might be to find segmentations of anatomical or pathological structures to track disease progressions \citep{cartiladge_measures_longitudinal}. For many segmentation tasks, not all details displayed in a fully-sampled MRI image are required, and larger structures are still recognizable even for high acceleration. The task of segmenting undersampled MRI images is therefore of high interest due to its applicability and its potential to save costly measurement time. 

\begin{figure*}
    \centering
    \includegraphics[width=0.9\linewidth]{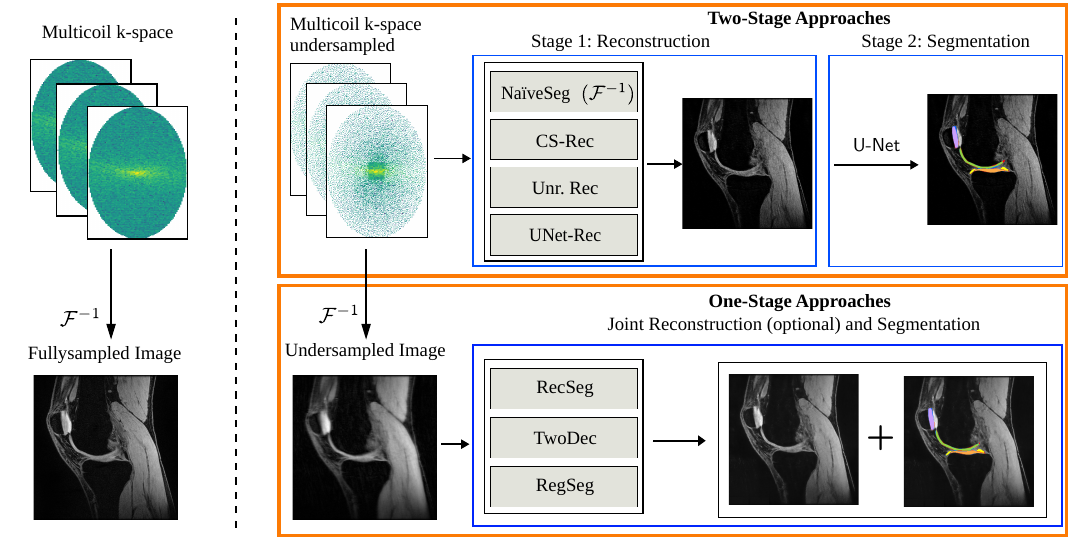}
        \caption{We separate modern methods that are optimized to segment undersampled MRI data into two categories: Two-Stage Approaches and One-Stage Approaches. Two-stage approaches first reconstruct the undersampled MR image and a simple UNet is then used to segment the reconstructed image. One-stage approaches try to combine both tasks and it is often claimed that this combination of tasks is helpful to achieve better reconstructions and segmentations. Our results suggest the contrary finding.}
    \label{fig:teaser_fig}
\end{figure*}

A number of works have explored methods to jointly optimize MRI reconstruction for optimal segmentation and reconstructions rather than for reconstruction fidelity alone \citep{RecSeg,k2s_challenge,pmlr-v121-caliva20a}.
Unfortunately, the majority of these methods suffer from common validation pitfalls \citep{isensee2024nnu} and do not compare to each other directly and instead only compare to simple baseline methods \citep{pmlr-v121-caliva20a,RecSeg}. Moreover, most methods only consider single and private datasets \citep{k2s_challenge,RecSeg,pmlr-v121-caliva20a} and sometimes only single acceleration factors \citep{k2s_challenge}, such that the behavior of the algorithms across datasets and across acceleration factors remains unknown.

To improve comparability between methods, a recent MICCAI challenge—the K2S Challenge \citep{k2s_challenge}—invited participants to perform segmentation given only undersampled k-space data and evaluated submissions based on the quality of the resulting segmentations. Interestingly, the winning method that predicted the best segmentations also obtained the lowest reconstruction fidelity of all submissions and did not explicitly optimize for high reconstruction scores. Moreover, the third-placed method skipped the reconstruction altogether and trained a segmentation network on the undersampled data directly. However, results from challenges such as the K2S Challenge should be interpreted with caution \citep{maier2018rankings,wiesenfarth2021methods}, as it is often unclear which specific factors within a method’s implementation are responsible for its performance \citep{Eisenmann_2023_CVPR}.
Apart from the network architecture, implementation decisions like data augmentation, number of iterations, and ensembling also play a crucial role and might be even more important than the underlying architecture \citep{nnunet}. Moreover, the K2S challenge was performed on a single and now private dataset. Therefore neither prior work nor the results of the K2S challenge can be used to answer the question of which method for segmenting undersampled MRI data performs the best and if high-quality reconstructions are indeed needed for better segmentation performance. 

In order to address this research gap and generate additional insight about segmenting undersampled MRI data, this paper provides the first fair comparison and a unified evaluation framework for many state-of-the-art methods in the current literature. By providing a rigorous benchmark and analysis for two datasets and several acceleration factors, we show that there is indeed no correlation between reconstruction scores in terms of PSNR and segmentation scores as measured by the Dice score \citep{dice1945measures}. Our results also show that most methods do not achieve statistically significant improvements compared to the simple baseline of segmenting the undersampled image, that is obtained through a Fourier transform from the undersampled k-space, directly. However, two-stage methods where the reconstruction methods enforce k-space consistency yield small yet still statistically significant better segmentation performance. This suggests that incorporating k-space consistency is a critical component for effective segmentation from undersampled MRI data.

\section{Related Work}

Only a small number of papers have focused on the task of obtaining segmentations directly from undersampled MRI data. In \citet{schlemper_2018}, the authors trained a cardiac-segmentation network on undersampled MRI-data directly, encouraging the creation of a consistent segmentation over different acceleration factors. In \citet{huang_brain_seg_unders} the authors used a combination of iterative reconstruction, an attention mechanism, a recurrent module, as well as a segmentation module to create adequate segmentations. They restricted their analysis, however, to one private single-coil dataset and did not analyze the effect of different acceleration factors on the segmentation quality. The authors of \citep{joint_rec_and_seg_ipmi_2018} combined the reconstruction with the segmentation task by using U-Net-like networks, in which the representation created by a shared encoder is used as input for the reconstruction and segmentation decoder. In a similar manner, \citet{pmlr-v121-caliva20a} have combined the MRI reconstruction and segmentation into one V-Net-like architecture that consists of a shared encoder and utilizes two decoders, one decoder for the reconstruction task and one for the segmentation task. In this method, the skip connection for the segmentation method is obtained from the reconstruction decoder instead of the encoder. Another line of work \citep{RecSeg, segmentation_aware_recon} indicates that good reconstruction and segmentation results can be achieved by making use of a reconstruction network in cascade with a segmentation network. The analyses, however, focus on the reconstruction scores and no in-depth analysis of effects on the Dice score is provided.

In a recent work by \citet{morshuis2024segmentation} the authors considered the uncertainty that results from the segmentation of undersampled MRI data. The authors analyzed the ill-posed nature of the MRI reconstruction problems: As there are infinite possible reconstructions that are data-consistent with the measured k-space, the authors generated several reconstructions corresponding to minimum and maximum segmentations. The true segmentation lies in-between these extreme cases. However, the authors did not try to maximize segmentation performance and instead focused on providing reliable uncertainty estimates.

Predicting high-quality segmentations has been the objective in the K2S challenge \citep{k2s_challenge}, where several methods for segmenting undersampled MRI data have been submitted. While reconstruction fidelity was also analyzed for the methods that did provide reconstructions, it was not used as an evaluation criterion. A drawback of the challenge is that only a single acceleration factor (8$\times$) was considered, leaving the behavior of the submitted algorithms on higher acceleration factors unknown. Moreover, submitted methods differed not only in architecture, but also in other crucial aspects of the training pipeline such as augmentation and optimization strategies, or ensembling of multiple networks. It is well-known that these factors can preclude any meaningful conclusions from challenge results \citep{wiesenfarth2021methods,maier2018rankings} and the success of a method can potentially be attributed to the wrong component of a system \citep{Eisenmann_2023_CVPR}. 

\section{Methods}
\label{sec:Methods_all}

\subsection{Data}
\label{sec:datasets}

In order to evaluate methods that create segmentations for undersampled MRI-data, datasets are required that provide both the raw k-space measurements and ground-truth segmentations. It is then possible to simulate accelerated MRI data by retrospectively masking the determined fraction of k-space points according to some undersampling pattern such as Poisson undersampling \citep{bridson2007fast}.

The K2S dataset \citep{k2s_challenge} consists of multi-coil k-space data of 300 3D isotropic MRI scans of the knee and segmentations of the bones and cartilage. One scan of the dataset was excluded in the following experiments, because the MRI-image provided has not been correctly aligned with the corresponding segmentation. From the remaining 299 images, a train-, val-, test-split of 240, 30, 29 has been created. An acceleration mask of 8 is provided with the challenge data. Masks of higher acceleration factors have been created using the \texttt{meddlr} library \citep{desai2021noise2recon} and the same Poisson undersampling pattern. The segmentation regions contained a total of 6 classes: The 3 bone-regions of the knee Patellar, Femoral and Tibial as well as the corresponding 3 cartilages.

Furthermore, we conducted experiments on the publicly available SKM-TEA dataset \citep{desai2021skm}. Similar to the K2S dataset mentioned above, this dataset consists of 155 3-dimensional multi-coil k-space data samples as well as their corresponding segmentations. Using the k-space data and the provided undersampled masks, we selected acceleration factors of 8$\times$ and 16$\times$, where 16$\times$ corresponds to the maximum acceleration factor for which an undersampling mask is provided by the dataset. The segmentation classes contain the patellar, femoral, and tibial cartilage as well as the meniscus. We use the official data split described in \citet{desai2021skm}, which results in a training, test, and validation set of 86, 33, and 36 images, respectively. 

The SKM-TEA MR images can be calculated from the k-space data by making use of 2D inverse Fast Fourier Transforms (iFFT), while the MRI images of the K2S data can only be calculated using a 3D iFFT. Image-voxels of the K2S images therefore depend on all measured k-space points of the MRI volume, while voxels in the SKM-TEA images only depend on the k-space points of the respective 2D MRI slice.

\null
\begin{figure*}[th]
    \centering
    \includesvg[width=0.89\linewidth]{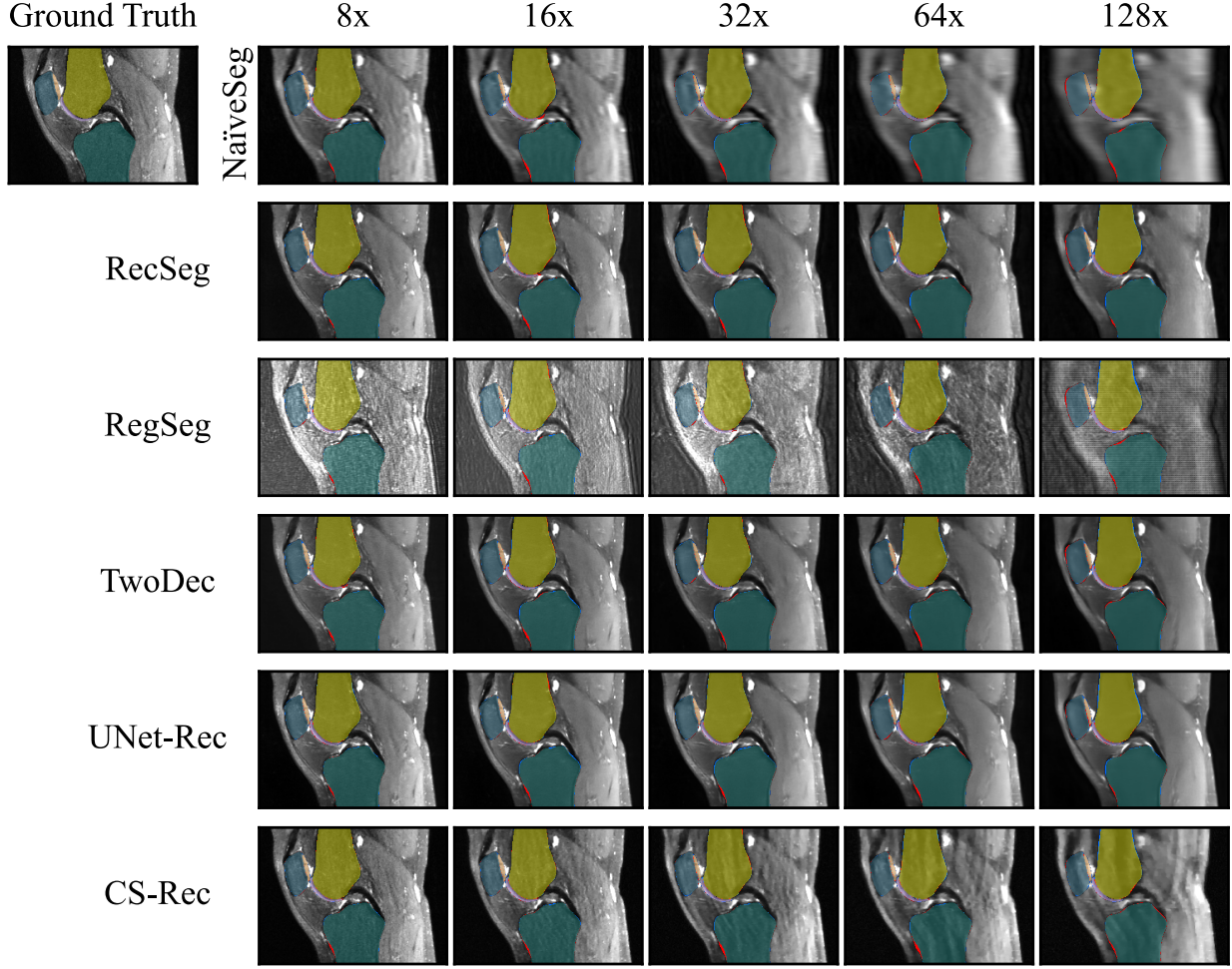}
    \caption{Example reconstructions and segmentations for the tested methods on the K2S dataset. False-positive segmentations are marked in red, and false negative segmentations are marked in blue. The RegSeg method does not necessarily generate reconstructions similar to the ground-truth, as this is no objective during training and instead the method focuses soley on the generation of good segmentations.}
    \label{fig:reconstructions}
\end{figure*}

\subsection{Reconstruction and Segmentation Methods}
\label{sec:segmentation_methods}
We analyze several state-of-the-art methods that have been developed for the segmentation of undersampled MRI data (see summary in Figure \ref{fig:teaser_fig}). We separate the methods into two-stage and one-stage methods: Two-stage methods first generate the reconstructions before predicting the segmentation in a second step, utilizing a standard U-Net segmentation network. The one-stage methods are specialized methods for segmenting undersampled MRI data and combine both tasks in a single forward pass.

\subsubsection{Two-stage methods}
\textbf{Na\"iveSeg}: This method relies on the creation of a zero-filled reconstruction by replacing the missing k-space measurements with zeros and applying an inverse Fourier transform. In a following step, we train a U-Net \citep{ronneberger2015u} based segmentation network directly on the zero-filled reconstructions. Because we utilize the method as a simple baseline method, we refer to this method as Na\"iveSeg. This approach was also submitted to the K2S challenge~\citep{naiveseg2023} where it reached third place. As for all the tested methods, we standardize the training framework and the networks' architecture which are both based on the nnU-Net framework \citep{nnunet} to improve comparability.

\textbf{UNet-Rec} \citep{hyun2018deep}: This two-stage approach utilizes an U-Net based reconstruction, originally proposed similarly by \citet{hyun2018deep}, yet we work with 3-dimensional (3D) reconstruction U-Nets with the same architectures as the ones used in nnU-Net \citep{nnunet}. First the images are reconstructed with a 3D U-Net. As reconstruction loss we use the  Mean-Squared-Error loss (MSE). After the reconstruction, we train the U-Net based segmentation network in the same manner as Na\"iveSeg on the reconstructed images.

\textbf{CS-Rec.} \citep{donoho2006compressed,lustig2007sparse}: In this approach, first a Compressed Sensing reconstruction (CS-Rec.) is applied to the undersampled image. CS-based reconstruction is a well-established method that is also applied in clinical practice \citep{donoho2006compressed,lustig2008compressed,candes2006robust}. We use a combination of Total Variation \citep{block2007undersampled} and L1-Wavelet \citep{candes2008enhancing} as the optimization objective. In a second step, a segmentation network is trained to segment the anatomical structures from the reconstructed image. 
We utilize the original code used by \cite{artem2023CS} the second-place submission in the K2S challenge \citep{k2s_challenge} for the image reconstruction. However, in contrast to the original method, we replace the V-Net segmentation architecture with the same U-Net architecture as Na\"iveSeg for better comparability to the other methods in our benchmark.

\textbf{Unr-Rec}
\citep{diamond2017unrolled}: Similarly to CS-Rec here we first generate reconstructions and then segment the image in a two-stage process. The reconstructions are obtained using the unrolled reconstruction method (Unr-Rec) proposed in \citep{diamond2017unrolled,sandino2020compressed} and also implemented as a baseline in the SKM-TEA dataset \citep{desai2021skm} that we utilize in our implementation. Unrolled reconstruction networks consist of multiple U-Nets in cascade with data-consistency elements in between. Due to computational constraints, unrolled reconstruction is currently limited to 2-dimensional imaging, as backpropagating through multiple 3D Fourier Transforms is computationally infeasible. The k-space structure of the SKM-TEA data allows for 2-dimensional Fourier transforms, but for the K2S data the full 3-dimensional k-space is required in order to calculate the inverse Fourier transform as mentioned in Section \ref{sec:datasets}. We therefore only evaluated this method on the SKM-TEA data. 

\begin{figure*}[t!]
    \centering
        \begin{subfigure}[t]{0.48\linewidth}
        \centering
    \includegraphics[width=0.95\linewidth]{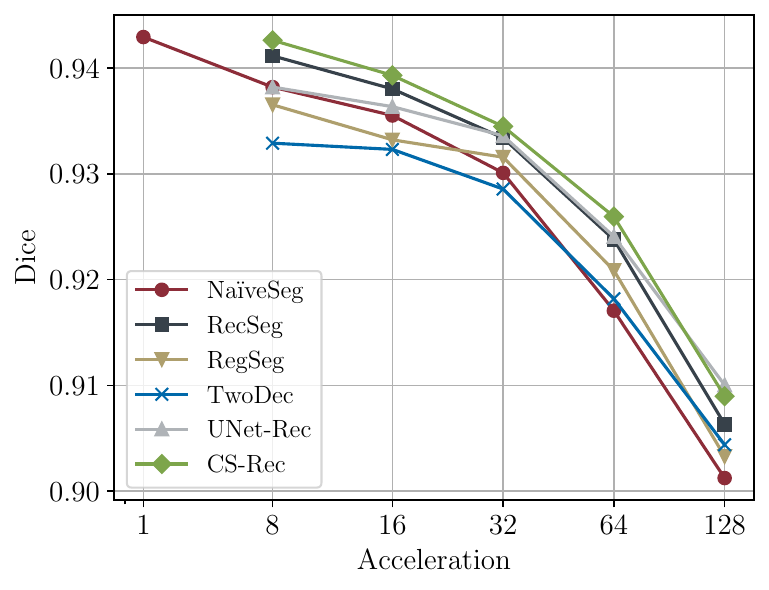}
    \end{subfigure}
    \hfill
    \begin{subfigure}[t]{0.48\linewidth}
        \centering
    \includegraphics[width=0.95\linewidth]{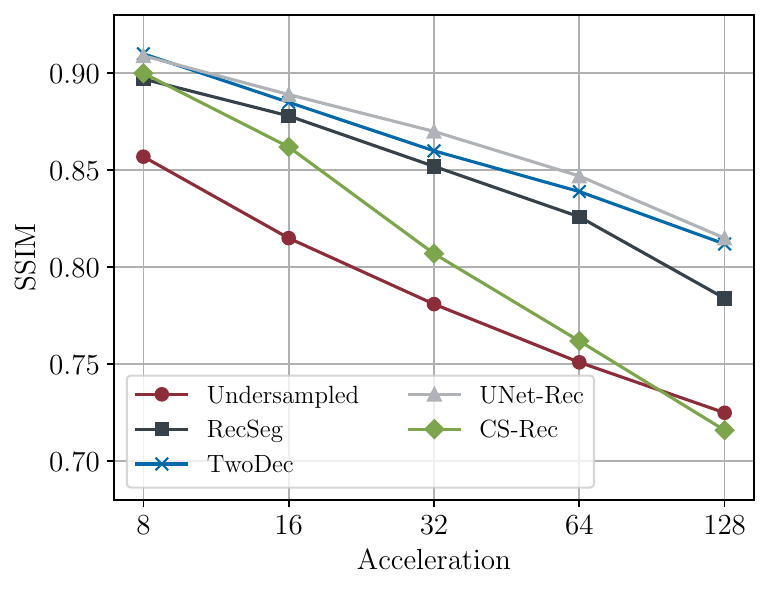}
    \end{subfigure}
    \caption{Comparison of the Dice score vs. the acceleration factor (left) as well as a comparison between the SSIM score and the acceleration factor (right) for the K2S dataset. Note that for better readability we did not include RegSeg scores in the SSIM plot on the right, as the method has solely been developed to improve segmentation scores and does not aim at creating a good reconstruction. We provide SSIM scores for RegSeg in Table \ref{tab:ssim_k2s}. Interestingly, CS-Rec achieves the highest  Dice scores while only achieving relatively low SSIM scores. The PSNR score behaves similarly as the SSIM score across accelerations.}
    \label{fig:segvsacc}
\end{figure*} 

\subsubsection{One-stage methods}

\textbf{RecSeg} \citep{RecSeg}: This method combines both the reconstruction and segmentation in a single forward pass. Specifically, the RecSeg model \citep{RecSeg} uses two U-Nets in a cascade, in which the first U-Net learns to reconstruct the MR image, while the second U-Net utilizes this reconstruction to estimate the segmentation. The difference to the UNet-Rec approach above is that here the two networks are trained simultaneously in an end-to-end fashion. This means that the reconstruction network also receives training gradients pushing it towards better segmentations. Following the original method we use an MSE-loss for training the reconstruction network, but use a Dice plus Cross-Entropy loss for the segmentation network as we also do for all other approaches.

\textbf{RegSeg} \citep{morshuis2023RegSeg}: This was the winning method of the K2S challenge \citep{k2s_challenge}. 
Similar to RecSeg, the RegSeg approach places a reconstruction and a segmentation network in a cascade. There are however two key differences to RecSeg, which is the optimization objective and the training method: RegSeg pre-trains both U-Nets independently for half the training epochs on the reconstruction and segmentation tasks by first using the fully-sampled MR image and the undersampled MR image as input for the segmentation network. After pre-training, the two U-Nets are placed in a cascade, where the segmentation network takes as input the reconstructed image as well as the zero-filled reconstruction of the original undersampled data. The two U-Nets are then trained for the second half of the epochs in an end-to-end manner on the segmentation task only. This allows the intermediate reconstructions to deviate from the fully-sampled image, thereby creating images that are potentially better suited for segmentation. These generated images will, however, potentially deviate strongly from the original reconstruction, as the generation of high-quality reconstructions is not the goal of RegSeg and instead the method focuses only on the segmentation quality.

\indent\textbf{TwoDec} \citep{pmlr-v121-caliva20a}: This method is an adaptation of the TB-Recon method originally proposed by \citet{pmlr-v121-caliva20a}. It consists of a shared encoder and two decoders that solve the reconstruction and the segmentation, respectively. While the original method employs a V-Net architecture, we opted for a U-Net-like base architecture. We also changed the loss function to include the Dice + Cross-Entropy loss as in all other segmentation models and adjusted the reconstruction function to MSE, as this loss is also utilized for UNet-Rec and RecSeg. This also allows us to isolate the key contribution of \citep{pmlr-v121-caliva20a} -- the one encoder, two decoder architecture -- and enables a fair comparison to the other methods. Moreover, we observed that a V-Net sometimes led to unstable training on the K2S dataset. As in the original publication we set the weight for the reconstruction loss and the segmentation loss to be equal for the K2S dataset and the SKM-TEA dataset at 8$\times$ acceleration, but do an hyperparameter search for the SKM-TEA dataset at 16$\times$ acceleration in Section \ref{sec:seg_quality}.

\subsection{Experimental Setup}

In MRI reconstruction, the goal is to find a reconstruction $\boldsymbol{\hat{x}}$ that approximates the ground-truth image $\boldsymbol{x}$ while remaining consistent with the measured k-space data $\boldsymbol{y}$: 
\begin{equation}
        \boldsymbol{y} = \boldsymbol{A}\boldsymbol{\hat{x}} \quad \text{ with }  \quad \bm{y} \in \mathbb{C}^m, \quad \bm{A} \in \mathbb{C}^{m \times n}, \quad \bm{\hat{x}} \in \mathbb{C}^n,
    \label{eq:inverse}
\end{equation}

where $\bm{A}=M\mathcal{F}S$ with the coil sensitivity maps $S$, the Fourier transform $\mathcal{F}$ and the undersampling mask $M$. 
In this work, our focus is not on predicting the most accurate reconstructions $\boldsymbol{\hat{x}}$ but on predicting high-quality segmentations $\boldsymbol{\hat{s}}$ that closely match the ground truth segmentations $\boldsymbol{s}$. In the \textbf{one-stage} approaches, the methods are applied directly on the MRI image that is derived from the zero-filled k-space data $\boldsymbol{y}$, such that the segmentation $\boldsymbol{\hat{s}}$ is predicted as $\bm{\hat{s}} = f_{\theta}(\boldsymbol{y})$. In the \textbf{two-stage} approaches the image is first reconstructed using existing reconstruction methods $\bm{\hat{x}} = g_{\psi}(\boldsymbol{y})$ and then segmented using a standard 3D U-Net framework ($f_{\phi}$) to estimate the segmentation $\bm{\hat{s}} = f_{\phi}(\bm{\hat{x}})$. We investigate which approach leads to the most accurate segmentations $\bm{\hat{s}}$ under potentially highly accelerated MRI acquisitions (see Fig. \ref{fig:teaser_fig} for a concept overview). Note that not all reconstruction methods require strict data-consistency as expressed in Equation \eqref{eq:inverse}.

The primary goal of this paper is to provide a rigorous benchmark of a representative set of recent methods that can segment undersampled MRI images. To enable meaningful comparisons, we standardize the evaluation setup to ensure maximal comparability across approaches. This, in turn, allows us to draw clear conclusions from our experiments and highlight promising directions for future research in this important yet underexplored field. 
In order to make the tested approaches maximally comparable, we modify the networks' architectures such that the employed components are the same and the differences in the score originate from changes in the tested method instead of the design of the network. 

To achieve comparability in the other aspects of the training setup like data-augmentation, training loss-function, data-preprocessing etc., we base all re-implemented methods on the well-established nnU-Net framework \citep{nnunet}, that was also utilized by the winning method of the K2S challenge \citep{k2s_challenge}. For all segmentation methods we therefore utilize a training loss consisting of a combination of the Dice loss and the Cross-Entropy loss. An high-level overview of similarities and differences is provided in Figure \ref{fig:teaser_fig}.
\subsection{Metrics}

We apply the Dice score to evaluate the predicted segmentations and utilize the SSIM \citep{ssim} and PSNR score in order to evaluate the quality of the predicted reconstructions. The metrics have been calculated using implementations provided by the MONAI framework \citep{monai}.

\subsection{Implementation Details}

We train and test all methods described in Section \ref{sec:segmentation_methods} from scratch on both the K2S and SKM-TEA datasets for all tested acceleration factors using a comparable training pipeline. All segmentation models are trained for 1000 epochs of 250 iterations on an Nvidia  Geforce RTX 2080Ti GPU using a batch-size of 2 and the Adam optimizer \citep{kingma2017adammethodstochasticoptimization}. We use a learning rate of $1e-2$ for most methods and $1e-3$ for RecSeg and TwoDec for training stability. The data-augmentation strategy is identical for all methods and consists of mirroring in the sagittal direction.

\section{Experiments and Results}
\label{sec:results}

\subsection{Segmentation Quality}
\label{sec:seg_quality}

The segmentation scores for all methods and all acceleration factors for the K2S dataset are shown in Figure~\ref{fig:segvsacc} (left). The quantitative values for the K2S, and the SKM-TEA datasets are further reported in Tables~\ref{tab:dice_k2s} and \ref{tab:skm}, respectively. We observed that, overall, the tested methods predicted segmentations of comparable quality and reached similar  Dice scores. For both datasets, the difference in  Dice score is within one standard deviation between the best and the worst performing method for all tested acceleration factors. The segmentations shown in Figure~\ref{fig:reconstructions} for the K2S dataset and in Figure~\ref{fig:recon_and_seg} for the SKM-TEA dataset show examples of the similar segmentation quality between the different methods. It can be seen that the differences in the segmentations are rather small within the same acceleration factor for all tested methods. We further note that the segmentations for all examined methods are of high quality and the errors are often of a similar nature for all methods, like the too large segmentation of the tibia bone as shown in Figure \ref{fig:reconstructions}.

In order to better understand the differences between the examined methods we analyzed them using a statistical significance test. Specifically, we compared the Dice scores of the tested models to the ones obtained by the simplest baseline method Na\"iveSeg using a Wilcoxon-Signed-Rank test~\citep{wilcoxon1992individual}. Statistically significant improvements in Dice score ($p<0.01$) are indicated by a star '$^*$' in Tables \ref{tab:dice_k2s} and \ref{tab:skm}. Even though the models are conceptually different, the Dice scores are rather similar despite varying SSIM scores of the reconstruction. Surprisingly, the CS-Rec method leads to the segmentation scores that are among the best for both datasets over nearly all acceleration factors. Furthermore, the results for CS-Rec are nearly always significantly better compared to Na\"iveSeg. Another surprising result is the good segmentation quality of Na\"iveSeg compared to the other more specialized methods.

\begin{figure*}[t]
    \centering
    \begin{subfigure}{0.48\linewidth}
            \includegraphics[width=0.98\linewidth]{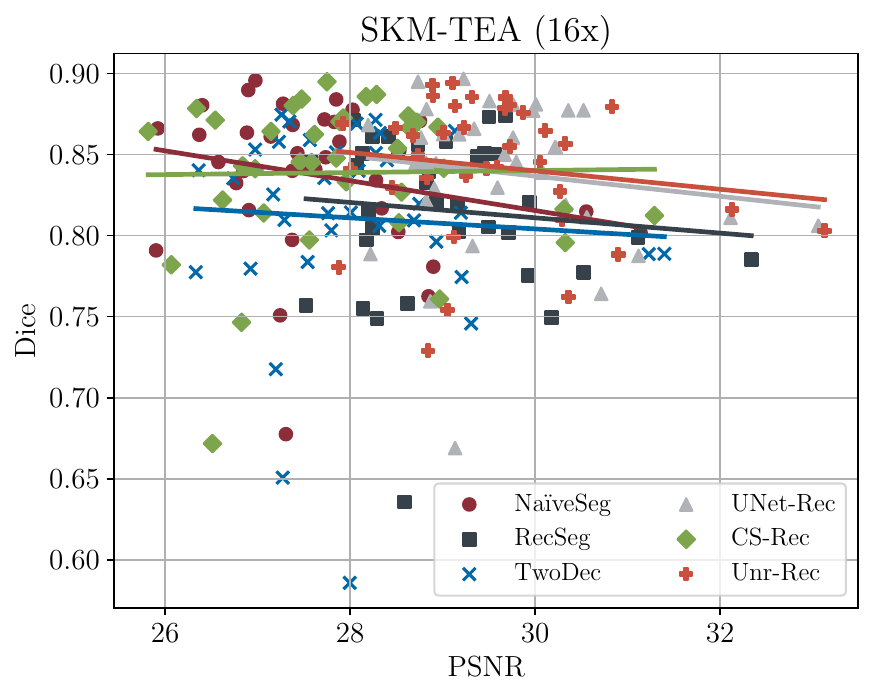}
    \end{subfigure}
    \begin{subfigure}{0.48\linewidth}
    \includegraphics[width=0.98\linewidth]{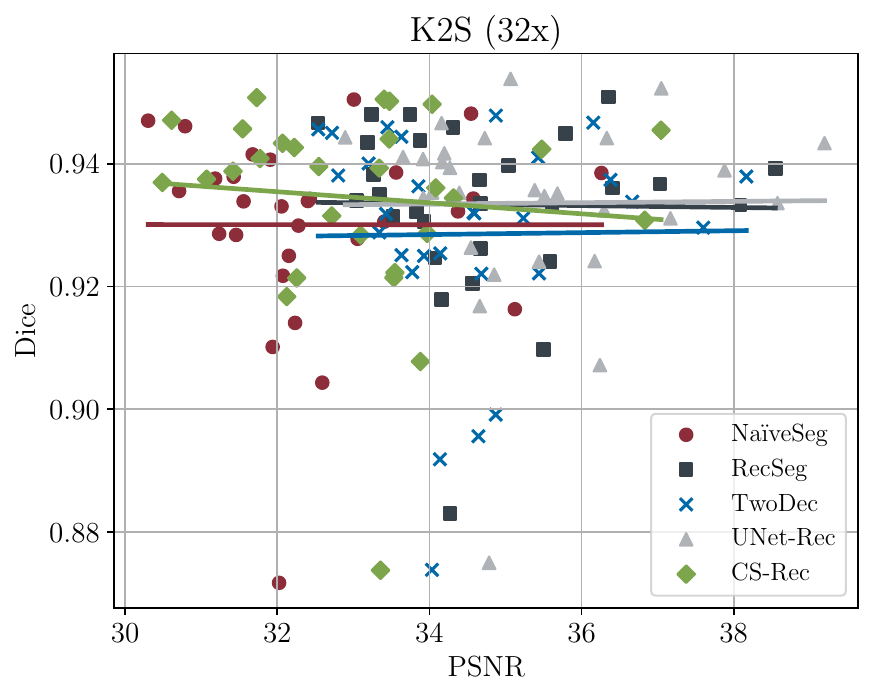}
    \end{subfigure}
    \caption{Relation between PSNR values and the Dice score for the SKM-TEA dataset with 16$\times$ acceleration (left) and the K2S dataset with 32$\times$ acceleration (right). Every reconstructed image on the test set is a single data point. It can be seen by the regression fit that no positive correlation exists between the PSNR and Dice scores within the methods. This observation is confirmed quantitatively with Spearman's $\rho$ ranging from $0$ to $-0.20$ (all $p-values>0.2$).}
    \label{fig:correlation-skm}
\end{figure*}

The method achieving the best Dice scores for the SKM-TEA dataset is Unr-Rec, which achieves the best scores for both tested acceleration factors and is the only method achieving statistically significantly better segmentation results at 16$\times$ acceleration. Note that Unr-Rec cannot be tested on the K2S data, because of the 3-dimensional k-space of the MRI images, as explained in Section \ref{sec:datasets}.

\begin{figure}[t]
    \centering
    \includegraphics[width=0.99\linewidth]{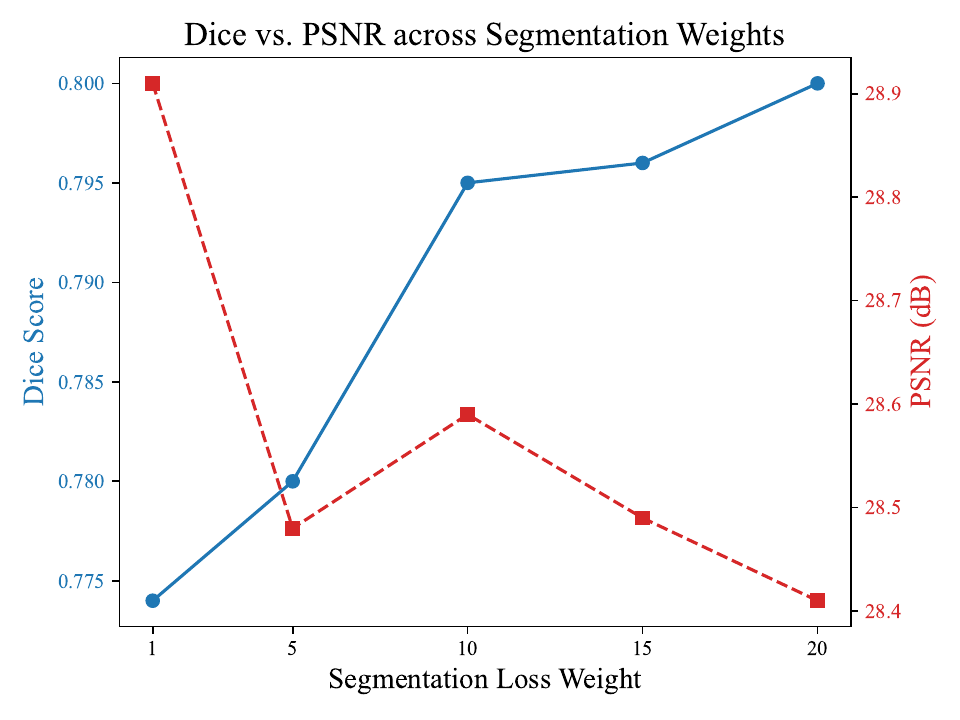}
    \caption{Analysis of segmentation weight for the TwoDec method on the validation set of SKM-TEA 16x data. Higher weights for the weight of the segmentation loss lead to higher segmentation scores, but the reconstruction quality also declines.}
    \label{fig:hyperparameter_twodec}
\end{figure}

We found that the performance of methods that combine the reconstruction and segmentation tasks like RecSeg and TwoDec can be dependent on the dataset and the acceleration factor. We observed that both methods are comparable to other methods on the K2S dataset, yet perform worse than other methods on the SKM-TEA dataset for 16$\times$ acceleration (see Table \ref{tab:skm}). We have analyzed the behavior of the TwoDec method on SKM-TEA with 16$\times$ acceleration when using higher weight (higher $\lambda_{seg}$ for the segmentation loss on the validation set
\begin{equation}
    \mathcal{L}_{total} = \mathcal{L}_{rec} + \lambda_{seg}\mathcal{L}_{seg}
\end{equation}
The results are shown in Figure \ref{fig:hyperparameter_twodec}. Note that higher segmentation weight leads to higher  Dice scores, but the reconstruction scores degrade. For the results in Table \ref{tab:skm} we chose $\lambda_{seg} = 10$, as it provides a good tradeoff between segmentation and reconstruction performance for the SKM-TEA dataset at 16$\times$ acceleration. Note that even though we have optimized the segmentation weight hyperparameter for the TwoDec method on the 16$\times$ acceleration experiments on SKM-TEA and thereby providing the TwoDec method with an advantage, the results are still worse compared to the other methods. 

\subsection{Reconstruction Quality}

In contrast to the  Dice scores, the reconstruction scores vary substantially between the methods. Interestingly, even though the segmentation scores are good for CS-based reconstructions, CS-Rec does not achieve very high SSIM and PSNR scores in both datasets. In Figure \ref{fig:segvsacc} the SSIM score is shown for several acceleration factors. It can be seen that all methods that aim to generate a reconstruction of the MR image improve SSIM for nearly all acceleration factors compared to the zero-filled reconstruction used in the method Na\"iveSeg. The reconstruction scores of CS-Rec decline more for higher accelerations compared to the other methods, even leading to lower reconstruction scores than the original zero-filled reconstruction. 

As can be seen in Figure \ref{fig:correlation-skm}, no correlation exists between the reconstruction score in terms of PSNR and the segmentation score when considering all predicted images and segmentations from the different methods individually. This indicates that the segmentation does not benefit from reconstructions that achieve higher PSNR values.

\begin{table*}[t]
    \centering
    \caption{Dice scores for the K2S dataset. Values represent the Mean $\pm$ Standard Deviation. The fully-sampled Na\"iveSeg baseline (1$\times$ acceleration) achieved $0.943\pm0.013$. Models with a statistically significantly ($p<0.01$) higher Dice score compared to the Na\"iveSeg model at the respective acceleration factor are marked with an asterix ($^*$). 
    The two-stage approach CS-Rec generally achieves the highest Dice scores. The scores remain stable up to 32$\times$ acceleration.}
    \label{tab:dice_k2s}
    \begin{tabular}{l c c c c c c }
    \toprule
        ~ & \multicolumn{6}{c}{Dice} \\ \cmidrule{2-7}
        ~ & 1$\times$ & 8$\times$ & 16$\times$ & 32$\times$ & 64$\times$ & 128$\times$ \\ \midrule
Na\"iveSeg     & 0.943 $\pm$ 0.013\phantom{$^*$}   &  0.938 $\pm$ 0.018\phantom{$^*$}  &  0.936 $\pm$ 0.015\phantom{$^*$}  &   0.930 $\pm$ 0.015\phantom{$^*$}  &   0.917 $\pm$ 0.020\phantom{$^*$}  &  0.901 $\pm$ 0.022\phantom{$^*$}  \\
RecSeg       & -   &  0.941 $\pm$ 0.012\phantom{$^*$}  &  0.938 $\pm$ 0.013\phantom{$^*$}  &  0.933 $\pm$ 0.013$^*$ &  0.924 $\pm$ 0.018$^*$ &  0.906 $\pm$ 0.024$^*$ \\
RegSeg       & -   &  0.937 $\pm$ 0.016\phantom{$^*$}  &   0.933 $\pm$ 0.020\phantom{$^*$}  &  0.932 $\pm$ 0.016\phantom{$^*$}  &   0.921 $\pm$ 0.020$^*$ &  0.903 $\pm$ 0.022\phantom{$^*$}  \\
TwoDec     & -   &  0.933 $\pm$ 0.021\phantom{$^*$}  &  0.932 $\pm$ 0.019\phantom{$^*$}  &  0.929 $\pm$ 0.018\phantom{$^*$}  &  0.918 $\pm$ 0.019\phantom{$^*$}  &  0.904 $\pm$ 0.024$^*$ \\
UNet-Rec     & -   &  0.938 $\pm$ 0.014\phantom{$^*$}  &   0.936 $\pm$ 0.020\phantom{$^*$}  &  \textbf{0.934 $\pm$ 0.015}$^*$ &  0.924 $\pm$ 0.016$^*$ &   \textbf{0.910 $\pm$ 0.019}$^*$ \\
CS-Rec & - &  \textbf{0.943 $\pm$ 0.012}$^*$ &  \textbf{0.939 $\pm$ 0.013}$^*$ &  \textbf{0.934 $\pm$ 0.016}$^*$ &  \textbf{0.926 $\pm$ 0.017}$^*$ &  0.909 $\pm$ 0.025$^*$ \\
        \bottomrule 
    \end{tabular}
\end{table*}

\begin{table*}[t]
    \centering
    \caption{SSIM scores for the K2S data. The reconstructions generated by the UNet-Rec methods usually achieve the highest SSIM scores. Note that the low scores of RegSeg can be explained by the fact that RegSeg does not aim to create a quality reconstruction and instead focuses on the segmentation.}
    \label{tab:k2s_ssim}
    \begin{tabular}{l c c c c c }
    \toprule
        ~ & \multicolumn{5}{c}{SSIM} \\ \cmidrule{2-6}
        ~ & 8$\times$ & 16$\times$ & 32$\times$ & 64$\times$ & 128$\times$ \\ \midrule
        Na\"iveSeg & 0.857 $\pm$ 0.024  & 0.815 $\pm$ 0.029  & 0.781 $\pm$ 0.034  & 0.751 $\pm$ 0.037  & 0.725 $\pm$ 0.039  \\ 
        RecSeg   &  0.897 $\pm$ 0.020  & 0.878 $\pm$ 0.023  & 0.852 $\pm$ 0.027  & 0.826 $\pm$ 0.032  & 0.784 $\pm$ 0.037  \\
        RegSeg   &  0.477 $\pm$ 0.074  &  0.603 $\pm$ 0.063  &  0.511 $\pm$ 0.076  &   0.610 $\pm$ 0.059  &  0.405 $\pm$ 0.071  \\
        TwoDec &   \textbf{0.910 $\pm$ 0.018} &  0.885 $\pm$ 0.021  &   0.860 $\pm$ 0.027  &  0.839 $\pm$ 0.031  &  0.812 $\pm$ 0.036  \\
        UNet-Rec & 0.909 $\pm$ 0.018  & \textbf{0.889 $\pm$ 0.022} & 
        \textbf{0.870 $\pm$ 0.025} & \textbf{0.847 $\pm$ 0.031} & \textbf{0.815 $\pm$ 0.033} \\
        CS-Rec & 0.900 $\pm$ 0.019  & 0.862 $\pm$ 0.027  & 0.807 $\pm$ 0.038  & 0.762 $\pm$ 0.044  & 0.716 $\pm$ 0.048  \\ 
        \bottomrule
    \end{tabular}
    \label{tab:ssim_k2s}
\end{table*}

\begin{figure*}[t!]
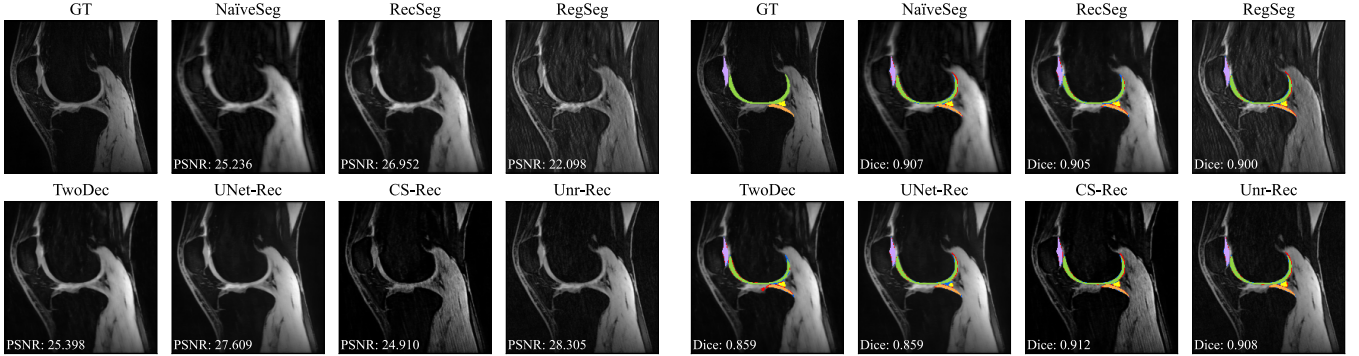

    \centering
    \begin{subfigure}{0.49\linewidth}
        \centering
        \includesvg[width=\linewidth]{plots/reconstruction_skm_rec.svg}
    \end{subfigure}
    \begin{subfigure}{0.49\linewidth}
        \centering
        \includesvg[width=\linewidth]{plots/reconstruction_skm_seg.svg}
    \end{subfigure}
    \caption{Qualitative results from the SKM-TEA dataset showing both reconstructions (left) and segmentations (right) at 16$\times$ acceleration. The segmentations show the following classes: Patellar Cartilage ({\color{patellar}$\blacksquare$}), Femoral Cartilage ({\color{femoral}$\blacksquare$}), Tibial Cartilage ({\color{tibial}$\blacksquare$}), Meniscus ({\color{meniscus}$\blacksquare$}), False Positives ({\color{fp}$\blacksquare$}), False Negatives ({\color{fn}$\blacksquare$}). RegSeg does not aim to generate perfect reconstructions but rather generates images helpful for the downstream segmentation task.}
    \label{fig:recon_and_seg}
\end{figure*}

\begin{table*}[t!]
    \centering
    \caption{PSNR scores for the K2S data. The Reconstruction provided by the UNet-Rec achieve the highest PSNR scores.}
    \label{tab:k2s_psnr}
    \begin{tabular}{l c c c c c}
    \toprule
            ~ & \multicolumn{5}{c}{PSNR} \\ \cmidrule{2-6}
        ~ & 8$\times$ & 16$\times$ & 32$\times$ & 64$\times$ & 128$\times$ \\ \midrule
        Na\"iveSeg & 35.0 $\pm$ 1.4 & 33.7 $\pm$ 1.4 & 32.5 $\pm$ 1.4 & 31.2 $\pm$ 1.4 & 30.2 $\pm$ 1.4 \\
        RecSeg & 37.1 $\pm$ 1.5 & 36.0 $\pm$ 1.5 & 34.7 $\pm$ 1.5 & 33.5 $\pm$ 1.5 & 32.0 $\pm$ 1.4 \\
        RegSeg & 26.0 $\pm$ 1.7 & 28.4 $\pm$ 1.6 & 27.4 $\pm$ 1.7 & 27.1 $\pm$ 1.7 & 25.3 $\pm$ 1.6 \\
        TwoDec & 36.7 $\pm$ 1.4 & 35.7 $\pm$ 1.4 & 34.6 $\pm$ 1.4 & 33.2 $\pm$ 1.4 & 32.3 $\pm$ 1.4 \\
        UNet-Rec & \textbf{37.6 $\pm$ 1.6} & \textbf{36.5 $\pm$ 1.5} & \textbf{35.4 $\pm$ 1.5} & \textbf{34.2 $\pm$ 1.5} & \textbf{32.8 $\pm$ 1.4} \\
        CS-Rec & 37.5 $\pm$ 1.6 & 35.5 $\pm$ 1.5 & 33.1 $\pm$ 1.6 & 31.3 $\pm$ 1.6 & 29.5 $\pm$ 1.5 \\
        \bottomrule
    \end{tabular}
\end{table*}

\begin{table*}[t!]
    \centering
    \caption{Scores for the SKM-TEA data. While Unr-Rec achieves the highest  Dice scores in our experiments, UNet-Rec achieves the highest SSIM and PSNR scores.}
    \begin{tabular}{l c c c @{\hskip 12pt} c c @{\hskip 12pt} c c}
    \toprule
    ~ & \multicolumn{3}{c@{\hskip 12pt}}{Dice} & \multicolumn{2}{c@{\hskip 12pt}}{SSIM} & \multicolumn{2}{c}{PSNR} \\
    \cmidrule(lr){2-4} \cmidrule(lr){5-6} \cmidrule(lr){7-8}
        ~ & 1$\times$ & 8$\times$ & 16$\times$ & 8$\times$ & 16$\times$ & 8$\times$ & 16$\times$ \\
        \midrule
Na\"iveSeg  & 0.854 $\pm$ 0.053&0.843 $\pm$ 0.044\phantom{$^*$} & 0.837 $\pm$ 0.045\phantom{$^*$}  &     0.736 $\pm$ 0.024 & 0.692 $\pm$ 0.026 & 28.6 $\pm$ 1.1  & 27.6 $\pm$ 1.1 \\
RecSeg         & - & 0.845 $\pm$ 0.047$^*$& 0.816 $\pm$ 0.049\phantom{$^*$} & 0.799 $\pm$ 0.024 & 0.739 $\pm$ 0.029 & 30.9 $\pm$ 1.0 & 29.4 $\pm$ 1.0 \\
RegSeg          & - & 0.847 $\pm$ 0.042$^*$& 0.838 $\pm$ 0.044\phantom{$^*$} & 0.582 $\pm$ 0.036 & 0.509 $\pm$ 0.040 & 25.5 $\pm$ 1.4 & 24.8 $\pm$ 1.3 \\
TwoDec        & - & 0.842 $\pm$ 0.043\phantom{$^*$} & 0.811 $\pm$ 0.060\phantom{$^*$} & 0.817 $\pm$ 0.021\phantom{$^*$} & 0.737 $\pm$ 0.026 & 30.8 $\pm$ 1.1 & 28.1 $\pm$ 1.1 \\
UNet-Rec        & - & 0.845 $\pm$ 0.044$^*$& 0.839 $\pm$ 0.045\phantom{$^*$} & \textbf{0.826 $\pm$ 0.024} & \textbf{0.750 $\pm$ 0.042} & \textbf{31.6 $\pm$ 1.0} & \textbf{29.7 $\pm$ 1.0} \\
CS-Rec & - & 0.847 $\pm$ 0.044$^*$& 0.839 $\pm$ 0.046\phantom{$^*$} & 0.680 $\pm$ 0.028 & 0.632 $\pm$ 0.030 & 29.0 $\pm$ 1.2 & 27.9 $\pm$ 1.2 \\
Unr-Rec          & - & \textbf{0.848 $\pm$ 0.045}$^*$ & \textbf{0.843 $\pm$ 0.041}$^*$ & 0.704 $\pm$ 0.035 & 0.645 $\pm$ 0.036 & 31.2 $\pm$ 1.1 &  29.5 $\pm$ 1.1  \\
\bottomrule
        \end{tabular}
        \label{tab:skm}
\end{table*}

Tables \ref{tab:ssim_k2s} and \ref{tab:skm} provide PSNR and SSIM scores for the reconstruction of different methods. The best reconstructions are created by the two-step UNet-Rec method for the majority of cases. As can be seen in Figures \ref{fig:reconstructions} and \ref{fig:recon_and_seg}, this method is also producing smooth results, which is known to be beneficial for the SSIM and PSNR metrics \citep{fastmri_challenge_2020}. Lower PSNR scores are reached with the SKM-TEA data in Table \ref{tab:skm} compared to the K2S data shown in Table \ref{tab:k2s_psnr}. An example image of the reconstructions and the segmentations for the SKM-TEA data with 16$\times$-acceleration factor is shown in Figure \ref{fig:recon_and_seg}.

\section{Discussion}
\label{sec:Discussion}

In our work, we have shown that, in contrast to previous findings \citep{k2s_challenge}, most methods for segmenting undersampled MRI data perform similarly for the same acceleration factors if the methods are trained in a comparable manner. Crucially, more complex approaches do not necessarily improve the segmentation score in a statistically significant way.

We observed that two-stage methods which first generate a reconstruction before predicting the segmentations generally perform better, both creating reconstructions leading to higher similarity scores as measured in PSNR and SSIM and also predicting more accurate segmentations. The best segmentation scores were obtained with the CS-Rec and Unr-Rec method that both (explicitly or implicitly) enforce data-consistency with the k-space data when predicting the reconstructed image.

The high segmentation quality obtained with CS-Rec is particularly interesting as the predicted reconstructions did not achieve high reconstruction scores compared to other approaches. The remaining methods that did not enforce k-space consistency often led to worse segmentation results despite having better reconstructions qualitatively and quantitatively in terms of SSIM (see for example the TwoDec approach). This suggests that reconstructions that are faithful to the measured k-space data may be more important for achieving good segmentations than reconstructions that obtain high visual fidelity. Thus, enforcing k-space consistency is likely to be an important feature for future work in this domain.

Another research question we investigated was whether high-quality reconstructions provide an advantage for accurate segmentation prediction. The missing correlation between reconstruction and segmentation scores shown in Figure \ref{fig:correlation-skm} suggests that a high-quality reconstruction is, indeed, not required to predict high-quality segmentations. Beyond CS-Rec, discussed above, two illustrative examples are RegSeg and Na\"iveSeg. Although neither method aims to predict a reconstructions, both achieve segmentation performance comparable to methods that explicitly optimize for image reconstruction, such as TwoDec or UNet-Rec. This suggests that future solutions aiming exclusively at obtaining high segmentation performance from highly accelerated k-space data do not necessarily need to additionally predict the reconstruction. However, we note that depending on the application high-quality reconstructions may be required for other reasons such as human interpretability. 

A disadvantage of the methods that combine the reconstruction with the segmentation task like RecSeg and TwoDec is the dependence on the dataset and acceleration factor. While the methods perform relatively well on the K2S Dataset, their performance degrades for the SKM-TEA dataset at accelerations of 16$\times$ as seen in Table \ref{tab:skm}. The TwoDec method is highly dependent on the selection of the loss weights. Careful tuning of those weights improved the segmentation score, yet the score was still worse than for the other methods.

In our experiments we have observed differences in the similarity scores of the predicted reconstructions on the two examined datasets. PSNR and SSIM values for the reconstructions on the SKM-TEA dataset were much lower compared to the K2S Dataset. This may indicate that the images of the SKM-TEA dataset were more difficult to reconstruct compared to the images of the K2S dataset. A possible reason for this could be that the SKM-TEA data is of higher resolution compared to the K2S data (512x512x160 vs 256x256x196) and can therefore display additional anatomical details that can make the reconstruction process difficult.

\section{Conclusions and Future work}

In this paper, we have shown that good segmentation quality of undersampled MRI images is feasible even for high acceleration of up to 32$\times$. Contrary to the findings in the K2S challenge \citep{k2s_challenge}, we have observed that reconstruction methods that enforce data-consistency - either strictly, as in CS-Rec, or implicitly, as in Unr-Rec - lead to the highest segmentation scores. However, we have also shown that the difference between the tested methods can be small, potentially making it difficult to see real-world benefits of the more complex and specialized methods. Experiments have also shown that the performance of some one-stage methods like RecSeg and TwoDec can depend on the dataset and acceleration factor. 

For future work that is looking at the segmentation of undersampled MR images, we suggest mainly two things: The task of reconstruction and segmentation can be treated independently, where first the reconstruction is generated and afterwards utilized as input for the 'na\"ive' UNet-based segmentation network as done in the tested two-stage methods. This leads to fewer hyperparameters that require tuning, to more stable results and to the best reconstruction (UNet-Rec) and segmentation scores (CS-Rec and Unr-Rec). Furthermore, we have seen that it is beneficial for the segmentation quality if the reconstruction method utilizes a data-consistency component, as is the case for CS-Rec and Unr-Rec.

Our research indicates that not all of the k-space information obtained in the fully-sampled image is required for an accurate segmentation prediction and high acceleration factors can be utilized. In the future, we believe it will be necessary to find a suitable trade-off between acceleration and segmentation quality that fits clinical standards while at the same time utilizes the MRI acquisition time more efficiently, allowing more patients to be treated and healthcare costs to be reduced.

\section*{Acknowledgements}
Funded by the Deutsche Forschungsgemeinschaft (DFG, German Research Foundation) under Germany’s Excellence Strategy - EXC number 2064/1
- Project number 390727645. The authors thank the International Max Planck Research
School for Intelligent Systems (IMPRS-IS) for supporting Jan Nikolas Morshuis.

\bibliographystyle{elsarticle-harv}
\bibliography{mybibliography}
\end{document}